\documentclass[12pt,english,floatfix,nofootinbib,superscriptaddress,aps,prd,preprint]{revtex4}
\usepackage[utf8]{inputenc}
\usepackage{float}
\usepackage{array}
\usepackage{lipsum}
\usepackage{graphicx}
\usepackage{amsmath,amsthm,amsfonts,amssymb}
\usepackage{graphicx}
\usepackage[english]{babel}
\usepackage{color}
\usepackage{tensor}
\usepackage{esint}
\usepackage[dvips]{epsfig}
\usepackage[dvips]{graphicx}
\usepackage{float}
\usepackage{units}
\usepackage{textcomp}
\usepackage{mathrsfs}
\usepackage{amsmath}
\usepackage[makeroom]{cancel}
\usepackage{amssymb}
\usepackage{amsbsy}
\usepackage{amsfonts}
\usepackage{amssymb,mathrsfs,xcolor}
\usepackage{esint}
\usepackage{array}
\usepackage{graphicx}
\usepackage{mathtools,slashed}
\usepackage{slashed}
\usepackage{hyperref}
\hypersetup{
    colorlinks,
    citecolor=blue,
    filecolor=green,
    linkcolor=purple,
    urlcolor=red,
}

\newcommand{\ie}{\begin{equation}}
\newcommand{\fe}{\end{equation}}
\newcommand{\se}{\begin{eqnarray}}
\newcommand{\ff}{\end{eqnarray}}

\usepackage{hyperref}
\hypersetup{colorlinks,breaklinks,
			citecolor=[rgb]{0,0.0,1.0},
            urlcolor=[rgb]{0.0,0.0,1.0},
            linkcolor=[rgb]{0,0.5,0.9}}

\begin{document}

\title{A novel nonlocal spin-1/2 theory: classical and quantum aspects.}

\author{J. R. Nascimento}
\email{jroberto@fisica.ufpb.br}
\affiliation{Departamento de Física, Universidade Federal da Paraíba, Caixa Postal 5008, 58051-970, João Pessoa, Paraíba,  Brazil} 

\author{Gonzalo J. Olmo}
\email{gonzalo.olmo@uv.es}
\affiliation{Departamento de F\'{i}sica Te\'{o}rica and IFIC, Centro Mixto Universidad de Valencia - CSIC. Universidad
de Valencia, Burjassot-46100, Valencia, Spain.}
\affiliation{Universidade Federal do Cear\'a (UFC), Departamento de F\'isica,\\ Campus do Pici, Fortaleza - CE, C.P. 6030, 60455-760 - Brazil.}

\author{A. Yu. Petrov}
\email{petrov@fisica.ufpb.br}
\affiliation{Departamento de Física, Universidade Federal da Paraíba, Caixa Postal 5008, 58051-970, João Pessoa, Paraíba,  Brazil} 

\author{P. J. Porfírio}
\email{pporfirio@fisica.ufpb.br}
\affiliation{Departamento de Física, Universidade Federal da Paraíba, Caixa Postal 5008, 58051-970, João Pessoa, Paraíba,  Brazil} 

\author{Ramires N. da Silva}
\email{rns2@academico.ufpb.br }
\affiliation{Departamento de Física, Universidade Federal da Paraíba, Caixa Postal 5008, 58051-970, João Pessoa, Paraíba,  Brazil} 

%%%%%%%%%%%%%%%%%%%%%%%%%%%%%%%%%%%%%%%%%%%%%%%%%%%%%%%%%%%%%%%%%%%%

\date{\today}

\begin{abstract}

We propose a novel nonlocal spin-1/2 theory in which the form factor depends on the Dirac operator rather than on the d'Alembert operator. In this scenario, we explore some classical and quantum aspects of this new theory. At the classical level, we investigate the dispersion relation of free spin-1/2 particles and find that it increasingly deviates from the standard case as the nonlocal effects become relevant. At the quantum level, we compute the fermionic one-loop effective action for the nonlocal spin-1/2 theory with Yukawa coupling and show that the contributions of nonlocal effects are significant in the UV limit, while in the IR they are suppressed by a UV cutoff scale, which has been chosen to coincide with the nonlocality scale $\Lambda$. We minimally couple a $U(1)$ gauge field to the nonlocal spin-1/2 field theory and explicitly demonstrate that this theory is gauge invariant. Finally, we obtain a nonlocal version of the Pauli equation and the impact of the nonlocality in the $g$-factor of massive particles.

\end{abstract}

\maketitle

\section{INTRODUCTION}

Nonlocal theories (theories that contain infinite-order derivative operators in their action) arise in different contexts in high-energy physics. The first works incorporating nonlocal effects in field theory were published a long time ago \cite{Wataghin,Pais,Pauli,Hayashi}. Further, Efimov \cite{Efimov} developed an ultraviolet (UV)-finite nonlocal extension of a quantum scalar field theory, specifically the S-matrix of a nonlocal scalar field theory. Nonlocal extensions of finite (non-)abelian gauge theories have also been constructed in \cite{Krasnikov, Moffat}. The Higgs mechanism was investigated within nonlocal field theories in \cite{Modesto4,Hashi}. In particular, the one-loop corrections were addressed in a nonlocal generalization of the scalar QED model \cite{Gama}. In the context of string/M-theories, nonlocal operators naturally emerge in the low-energy limit of these theories \cite{Kaku, Green, Witten:1985}, as a result of $\alpha^{\prime}$-corrections ($\alpha^{\prime}$ is defined as the inverse of the string tension). Another example of a nonlocal model is given by $p$-adic string theory \cite{Freund, Freund2, Moeller}. In gravitational scenarios the situation is much more subtle since, as is well-known, general relativity (GR) is a non-renormalizable theory. However, the inclusion of higher-order curvature counterterms cancels out the divergences in the effective action, thereby improving their UV behavior \cite{Shapiro}. Conversely, the presence of higher-order curvature terms spoils the unitarity of the theory, leading inevitably to the appearance of negative-norm states, the so-called ghosts. Nonlocal extensions for gravity theories were originally proposed to resolve the problem of the appearance of ghosts and, at the same time, to provide renormalizability or even UV finiteness for a theory \cite{Tomboulis,Modesto1,Modesto2,Modesto3,Modesto5,Boos,Biswas:2011ar}. At the classical level, several studies of nonlocal gravity theories have been performed in different scenarios, for example, nonlocal cosmological models \cite{Biswas:2005qr,Biswas,Jhingan,Biswas2,Dimitrijevic,Modesto8,Koshelev,Koshelev2}, nonlocal black hole solutions \cite{Modesto6, Modesto7} and other exact solutions within nonlocal theories \cite{Buoninfante1,Buoninfante2,Andrade,Nascimento, Ghoshal:2022mnj}.

Perhaps the most appealing property of these theories is the avoidance of the emergence of unpleasant ghost-like degrees of freedom. This is usually achieved by replacing the d'Alembert operator $\Box$ with the nonlocal operator $\Box f(\Box)$, with $f(\Box)$ being an entire function (also known as form factor) of its argument. The simplest example could be taking $f(\Box)=e^{\pm\Box/\Lambda^2}$, where $\Lambda$ is some high-energy characteristic scale, its inclusion in the Lagrangian -- for instance, $\mathcal{L}=\frac{1}{2}\Phi(\Box e^{\pm\Box/\Lambda^2}-m^2)\Phi$ -- prevents the emergence of additional degrees of freedom  (for examples of quantum studies of such theories, see \cite{Briscese,Eugenio}). Similarly, nonlocal gauge theories, including supersymmetric ones \cite{Gama2,Gama3}, and gravitational theories (see the references above) have been constructed. Nevertheless, for reasons that we ignore, nonlocal spinor theories have received very little attention in the literature. Some attempts were made in \cite{Krasnikov,Moffat}. In both cases, the authors considered form factors as functions of the d'Alembert operator, which seems to be unnatural to describe a nonlocal spin-1/2 field theory, because the Dirac Lagrangian depends only on the first-order derivatives through the Dirac operator, $\slashed{\partial}=\gamma^{\mu}\partial_{\mu}$. For this reason, we believe that the most natural approach to construct a nonlocal spin-1/2 field theory should be by promoting the $\slashed{\partial}$ operator to the more general form $\slashed{\partial} f(\slashed{\partial})$, where $f(\slashed{\partial})$ is an entire function of the Dirac operator. Here we focus on the implementation of this approach in a nonlocal field theory for spinors. 

The structure of the paper is as follows: in Section II, we briefly describe the main properties of nonlocal bosonic field theories, highlighting the subtleties faced when dealing with fermionic fields. In Section III, we construct a new nonlocal spin-1/2 theory in Minkowski spacetime by promoting the Dirac operator $\slashed{\partial}$ to a nonlocal operator $\slashed{\partial} f(\slashed{\partial})$. We carefully derive its field equations and show that its local limit recovers the standard Dirac equation. In addition, we obtain the dispersion relation of free spin-1/2 particles within this nonlocal model. In Section IV, we focus on the quantum aspects. In particular, we add a Yukawa coupling to the nonlocal spin-1/2 theory and then compute the Euclidean fermionic one-loop effective action. Since the nonlocal loop integrals are challenging to compute exactly, we adopt an approximation method to proceed with these calculations. In Section V, we construct a $U(1)$ gauge-invariant nonlocal spin-1/2 field theory by minimally coupling an electromagnetic field with the nonlocal spinor field, and we obtain a nonlocal version of the Pauli equation and the impact of the nonlocality in the $g$-factor of massive particles.

\section{General setup}

 Higher-order derivative field theories have been explored in various contexts. However, these theories generally suffer from unavoidable Ostrogradsky-like instabilities at the classical level, resulting in a Hamiltonian that is not bounded from below \cite{Ostrogradsky}. At the quantum level, the particle spectrum includes ghost states—degrees of freedom characterized by propagators with ``wrong'' signs (negative-energy particles), which spoil the unitarity of the quantum theory \cite{Hawking,Antoniadis}. To circumvent these issues, a common approach is to employ nonlocal field theories, which naturally emerge in low-energy limit of string theory. Within these models, the concept of nonlocality is incorporated in the Lagrangian by including form factors containing infinite-order derivatives \cite{Witten:1985}. 

As a starting point let us consider the following  nonlocal massless scalar field theory
\begin{equation}
\mathcal{L}=\frac{1}{2}\Phi\Box f(\Box)\Phi,
\end{equation}
where the operator $\Box=\eta^{\mu\nu}\partial_{\mu}\partial_{\nu}$ denotes the d'Alembert operator, and $f(\Box)$ is called form factor, which is an entire function of $\Box$. To ensure the unitarity of the theory, the form factor must be an entire function of the d'Alembert operator to avoid the emergence of ghost states. In fact, as a consequence of the Weierstrass factorization theorem \cite{Complex}, $f(\Box)=e^{h(\Box)}$, where $h(\Box)$ is also an entire function. Thus, the propagator does not include any additional degrees of freedom in theory when compared with the local (standard) one,
\begin{equation}
    \langle\Phi(-p)\Phi(p)\rangle=-i\frac{e^{-h(-p^2)}}{p^2},
    \label{EQ2new}
\end{equation}
which $p$ is the usual 4-momentum, it shows the presence of a unique pole, $p=0$, as expected. In addition, the ultraviolet (UV) behavior of the theory can be improved by choosing a well-motivated entire function $h(\Box)$ without introducing additional degrees of freedom.

%However, it is important to note that this requirement alone is not enough to guarantee the renormalizability of the theory, as indicated in REF. Therefore, for the purposes of this work, we restrict the form factors to be entire functions.

  Formally, an entire function can be expressed as a power series in its argument. In particular, we have
\begin{equation}
f(\Box)=e^{h(\Box)}=\sum_{n=0}^{\infty}c_{n}\Box^{n}_{\Lambda},
\end{equation}
where \(c_{n}\) are the coefficients of the power series (with \(c_0 = 1\), to provide a correct IR limit), and we have defined \(\Box_{\Lambda} \equiv \Box/\Lambda^2\). Here, we have introduced \(\Lambda\) which represents a typical high-energy scale (such as the Planck scale or string scale) at which nonlocal effects become relevant. As one approaches the \(\Lambda\) scale, we mean \(\mathcal{O}(\Box/\Lambda^2) \to 1\), then the UV regime is achieved. Conversely, the infrared (IR) regime is attained when \(\Lambda \to \infty\) and/or when integrating over small momenta, \(k \to 0\).

We have so far outlined the main properties of a massless nonlocal scalar field theory. However, nonlocality has been introduced in other contexts, such as gauge and gravity theories. In particular, in the context of nonlocal gravity, different choices of form factors have been considered. One particularly well-motivated choice is given by \cite{Tomboulis}
\begin{equation}
    f(\Box)=\frac{e^{-H(\Box_{\Lambda})}-1}{2\Box_{\Lambda}},
\end{equation}
where 
\begin{equation}
    H(z)=\frac{a}{2}\left(\Gamma(0,p(z)^2)+\gamma_E +\log(p(z)^2)\right).
\end{equation}
In this expression, \(\gamma_E\) denotes the Euler-Mascheroni constant, \(\Gamma(0, p(z)^2)\) represents the incomplete gamma function, and \(p(z)\) is a polynomial that satisfies the initial condition \(p(0) = 0\). 

On the other hand, as far as we know, nonlocal modifications of the Dirac Lagrangian have not received much attention in the literature, except the nonlocal quantum electrodynamics and nonlocal electroweak theory proposed in \cite{Moffat}, where nonlocal extensions in the fermion sector were also considered. Here, we intend to make further progress in this direction by proposing a novel nonlocal spinor field theory.

\section{Free nonlocal spin-$1/2$ theory}\label{sec2}

Our aim now is to construct a consistent nonlocal Dirac spinor theory. First of all, it is important to highlight a key point that distinguishes spinor theories from other theories involving scalar fields or gravity. In spin-$1/2$ Dirac theory, the dynamical term is made up of the free Dirac operator, $\slashed{\partial}=\gamma^{\mu}\partial_{\mu}$. Thus, this operator is a first-order derivative operator. Nevertheless, the nonlocal spin-$1/2$ theories so far discussed in the literature present form factors built as functions of $\Box$ (see, for example, \cite{Moffat}). A more natural approach to implementing nonlocality in the fermionic sector should make use of the Dirac operator as a basic building block to define the corresponding form factors. Note that, as usual, the squaring of the Dirac operator can be expressed in terms of the d’Alembert operator by using the properties of the Dirac matrices, namely, $\slashed{\partial}^2=-\Box \hat{I}$. Here, the $(-)$ sign appears due to the metric signature choice $(-,+,+,+)$.  So, this is the main aim of our investigation.

Let us introduce a generic nonlocal free spin-$1/2$ Lagrangian in Minkowski spacetime, which is given by 
\begin{equation}   
\mathcal{L}_{o}=\bar\Psi[i\slashed{\partial} f(\slashed{\partial})-m]\Psi + h.c,
    \label{eq1}
\end{equation}
where $f(\slashed{\partial})$ is the form factor,
defined as an entire function of the Dirac operator $\slashed{\partial}$. By varying Eq.  \eqref{eq1} with respect to $\bar\Psi$, one gets the nonlocal Dirac equation
\begin{equation}
        [i\slashed{\partial} f(\slashed{\partial})-m]\Psi=0.
    \label{eq2}
\end{equation}
Following the same procedure as the standard (local) case, namely, ``squaring'' the former equation to obtain 
\begin{equation}
        [\slashed{\partial}^{2} f^{2}(\slashed{\partial})+m^{2}]\Psi=0,
    \label{klein}
\end{equation}
from which it is straightforward to obtain the corresponding propagator in the momentum space
\begin{eqnarray}
    \langle\Psi(-p)\Psi(p)\rangle=\frac{i}{\slashed{p}f(\slashed{p})-m}=\frac{i(\slashed{p} f(\slashed{p})+m)}{p^2 f^2(\slashed{p})-m^2}.
\end{eqnarray}
It is easy to see that since $f(\slashed{p})$ is an entire function, no additional poles appear in the propagator.
From the mathematical point of view, an entire function is an analytic function throughout the complex plane, and then it can be expanded as a power series of its argument, which is now a matrix. In our particular case, the form factor reads
\begin{equation}
        f(\slashed{\partial})=\sum
        \limits_{n=0}^{\infty} b_{n} \slashed{\partial}_{\Lambda}^{n},
    \label{f}
\end{equation}
where $\slashed{\partial}_{\Lambda}\equiv \dfrac{\slashed{\partial}}{\Lambda}$ is dimensionless, $b_{n}$ are the coefficients (dimensionless) of the power series, satisfying the condition $b_{0}=1$.

Consider now the spectrum of the square of the local Dirac operator:  
\begin{equation}
    \slashed{\partial}^2 \Psi=-\Box\Psi=\lambda^2 \Psi,
\end{equation}
where $\lambda^2$ is the eigenvalue of the Dirac operator squared. Applying the operator $\slashed{\partial}$ successively on the left-hand side of this equation, elementary algebraic manipulations lead to the relations
\begin{eqnarray}
\nonumber  \slashed{\partial}^3 \Psi&=&-\slashed{\partial}\Box\Psi=\lambda^2 \slashed{\partial}\Psi;\\
  \nonumber  \slashed{\partial}^4 \Psi&=&\lambda^2\slashed{\partial}^2\Psi=\lambda^4\Psi;\\
  \nonumber &.&\\
  \nonumber &.&\\
  \nonumber &.&\\
\label{2m}\slashed{\partial}^{2m+1}\Psi&=&\lambda^{2m}\slashed{\partial}\Psi,\,\,\,\,\,m \in \mathbb{N};\\
\label{2n}\slashed{\partial}^{2n}\Psi&=&\lambda^{2n}\Psi,\,\,\,\,\,n \in \mathbb{N}.
\end{eqnarray}

 From Eq.\eqref{f}, we note that
\begin{eqnarray}
\nonumber\slashed{\partial}f(\slashed{\partial})&=& \slashed{\partial}\left[\sum_{n=0}^{\infty}b_{n}\slashed{\partial}^n_{\Lambda} \right]\\
\nonumber&=&\slashed{\partial}\left[\sum_{n=0}^{\infty}b_{2n}\slashed{\partial}^{2n}_{\Lambda}+\sum_{m=0}^{\infty}b_{2m+1}\slashed{\partial}^{2m+1}_{\Lambda}\right]\\
\nonumber&=&\left[\sum_{m=0}^{\infty}b_{2m+1}({-1})^{m+1}\frac{\Box^{m+1}}{\Lambda^{2m+1}}\right]\hat{I}+\left[\sum_{n=0}^{\infty}b_{2n}({-1})^{n}\frac{\Box^n}{\Lambda^{2n}}\right]\slashed{\partial}\\
&=& f_{1}(\Box)\hat{I}+f_2(\Box)\slashed{\partial},\label{k}
\end{eqnarray}
 where, using Eqs. \eqref{2n} and \eqref{2m}, we check that
\begin{equation}
    f_{1}(\Box)\Psi=\sum_{m=0}^{\infty}b_{2m+1}\frac{\lambda^{2m}}{\Lambda^{2m+1}}\Psi
\end{equation}
 and 
 \begin{equation}
    f_{2}(\Box)\Psi=\sum_{n=0}^{\infty}b_{2n}\left(\frac{\lambda}{\Lambda}\right)^{2n}\Psi
\end{equation}
  are entire functions built out from the d'Alembert operator. The second term in the right-hand side of Eq.\eqref{k} coincides with the standard nonlocal contribution widely discussed in the literature as \cite{Moffat, Carone1}. By contrast, the first term in the right-hand side of Eq.\eqref{k} represents, as far as we know, a new nonlocal deformation in the kinetic sector of the Dirac action, without precedent in previous works. Thus, the nonlocal Dirac Lagrangian \eqref{eq1} can be split into two nonlocal pieces, i.e.,
  \begin{equation}
\mathcal{L}_{o}=\bar\Psi[i\slashed{\partial} f_2(\Box)-m]\Psi+i\bar{\Psi}f_1(\Box)\Psi + h.c.   
  \end{equation}
Similarly, the nonlocal Dirac equation \eqref{eq2} becomes
\begin{equation}
    \left[i\slashed{\partial} f_2(\Box)-m+if_1(\Box)\right]\Psi=0. 
\end{equation}

Now, by squaring Eq.\eqref{k} and using Eqs. \eqref{2n} and \eqref{2m}, it follows that
\begin{equation}
    \slashed{\partial}^{2} f^{2}(\slashed{\partial})\Psi=-\Box\, g(\Box,\slashed{\partial})\Psi,
\end{equation}
where the nonlocal operator $g(\Box,\slashed{\partial})$ can be expanded as
\begin{eqnarray}  g\left(\Box,\slashed{\partial}\right)\Psi=(a_1(\Box)\hat{I}+a_2(\Box)\slashed{\partial})\Psi,
    \label{gnon}
\end{eqnarray}
with 
\begin{equation}  a_1(\Box)\Psi=\sum_{k=0}^{\infty}c_{2k}\left(\frac{\lambda}{\Lambda}\right)^{2k}\Psi,
\label{a1}
\end{equation}
and
\begin{equation}
a_2(\Box)\Psi=\sum_{k=0}^{\infty}c_{2k+1}\frac{\lambda^{2k}}{\Lambda^{2k+1}}\Psi,
\label{a2}
\end{equation}
where the $c_{i_{s}}$ are the coefficients of the power series of $f^{2}(\slashed{\partial})$, with $c_{0}=1$. To have more explicit results, it is essential to choose a concrete representation of the form factor. In particular, let us consider two different proposals, namely,
\begin{eqnarray}
    \label{f1}f_{I}(\slashed{\partial})&=&e^{-\frac{\slashed{\partial}}{\Lambda}}\\
    \label{IIi}f_{II}(\slashed{\partial})&=&e^{-i\frac{\slashed{\partial}}{\Lambda}}
\end{eqnarray}
For the former, the series can be broken into two terms, i.e.,
\begin{equation}
        f_{I}(\slashed{\partial})=\sum
        \limits_{n=0}^{\infty}
        \frac{1}{2n!}\frac{\slashed{\partial}^{2n}}{\Lambda^{2n}}+\sum
        \limits_{m=0}^{\infty}
        \frac{-1}{(2m+1)!}\frac{\slashed{\partial}^{2m+1}}{\Lambda^{2m+1}}
    \label{fa}
\end{equation}

Plugging Eqs. \eqref{2n} and \eqref{2m} into Eq. \eqref{fa}, one finds
\begin{eqnarray}
       \nonumber f_{I}(\slashed{\partial})\Psi&=& \left(\sum \limits_{j=0}^{\infty} \frac{\lambda^{2j}}{(2j)!\Lambda^{2j}}
        \hat{I}-\frac{1}{\lambda} \sum \limits_{k=0}^{\infty} \frac{\lambda^{2k+1}}{(2k+1)!\Lambda^{2k+1}}\slashed{\partial}\right)\Psi\\
        &=& \left(\cosh{\left(\frac{\lambda}{\Lambda}\right)}\hat{I}-\frac{1}{\lambda}\sinh{\left(\frac{\lambda}{\Lambda}\right)}\slashed{\partial}\right)\Psi,
    \label{fa3}
\end{eqnarray}
and, as a consequence,
\begin{equation}
        f_{I}^{2}(\slashed{\partial})= \cosh{\left(\frac{2\lambda}{\Lambda}\right)}\hat{I}-\frac{1}{\lambda}\sinh{\left(\frac{2\lambda}{\Lambda}\right)}\slashed{\partial}.
    \label{f2a}
\end{equation}
Putting all this information together, the nonlocal spin-$1/2$ Dirac equation \eqref{klein} can be cast into a more compact form, namely,
\begin{equation}
         [\Box\, g(\Box, \slashed{\partial}) -m^{2}]\Psi=0,
    \label{nlde}
\end{equation}
where we have defined the nonlocal operator 
\begin{equation}
g(\Box, \slashed{\partial})\Psi= f_{I}^{2}(\slashed{\partial})\Psi=\left( \cosh{\left(\frac{2\lambda}{\Lambda}\right)}\hat{I}-\frac{1}{\lambda}\sinh{\left(\frac{2\lambda}{\Lambda}\right)}\slashed{\partial}\right)\Psi.
\label{kht}
\end{equation}
Comparing it with Eqs. \eqref{a1} and \eqref{a2}, it is straightforward to see that
\begin{eqnarray}
\label{a3}    a_1(\Box)\Psi&=&\cosh\left(\frac{2\lambda}{\Lambda}\right)\Psi,\\
    a_2(\Box)\Psi&=&-\frac{1}{\lambda}\sinh\left(\frac{2\lambda}{\Lambda}\right)\Psi,\label{a4}
\end{eqnarray}
which are non-polynomial functions.
Proceeding similarly, we can find that the second form factor \eqref{IIi} satisfies
\begin{equation}
    g(\Box, \slashed{\partial})\Psi= f_{II}^{2}(\slashed{\partial})\Psi=\left( \cos{\left(\frac{2\lambda}{\Lambda}\right)}\hat{I}-\frac{i}{\lambda}\sin{\left(\frac{2\lambda}{\Lambda}\right)}\slashed{\partial}\right)\Psi.
    \label{kht2}
\end{equation}

It is worth calling attention to the fact that  Eq. \eqref{nlde} is not a nonlocal Klein-Gordon equation, as in the local case\footnote{The local case is recovered by taking $\Lambda \to \infty$. In this limit, the term proportional to the Dirac operator vanishes.}. This happens because of the presence of the Dirac operator $\slashed{\partial}$ on the right-hand side of Eqs. \eqref{kht} and \eqref{kht2}. Therefore, each spinor component must fulfill a more complex equation than the nonlocal Klein-Gordon one.

\subsection{Dispersion relation}

Expressing the solution to Eq. \eqref{nlde} as given by a superposition of plane waves, 
\begin{equation}
    \Psi(x)=\int d^{4}p \,e^{-ip_{\mu}x^{\mu}}\tilde{\Psi}(p),
\end{equation}
where $\tilde{\Psi}(p)$ is the Fourier transform of the spinor $\Psi(x)$, it must satisfy the nonlocal Dirac equation
\begin{equation}
  [p^2g(p^2,\slashed{p})+m^2\big]\tilde{\Psi}=0,
\end{equation}
where $g(p^2,\slashed{p})$ is the Fourier transform of the nonlocal operators \eqref{kht} and \eqref{kht2}. Non-trivial solutions of the above equation are obtained by requiring that the determinant of the matrix $S\equiv p^2g(p^2,\slashed{p})+m^2 \hat{I}$ satisfies the condition
\begin{equation}
    \det S=0, 
\end{equation}
 which provides the dispersion relation. To find it explicitly, let us proceed with some algebraic manipulations. Note that, rewriting the form factor in terms of the nonlocal operators \eqref{gnon}, in the momentum space, we have 
 \begin{eqnarray}
\nonumber \det S&=&\det\left[\left(p^2 a_1(p)+m^2\right)\hat{I}+p^2 a_2(p)(i\slashed{p})\right]\\
      &=& \det\bigg[\left(p^2 a_1(p)+m^2\right)\hat{I}\bigg]\times\det\left[\hat{I}+\frac{p^2 a_2(p)(i\slashed{p})}{p^2 a_1(p)+m^2}\right].
      \label{det}
 \end{eqnarray}
 The first determinant on the right-hand side of the above equation is easily obtained,
 \begin{equation}
     \det\bigg[\left(p^2 a_1(p)+m^2\right)\hat{I}\bigg]=\left(p^2 a_1(p)+m^2\right)^4,
     \label{det1}
 \end{equation}
 where $a_1(p)$ and $a_2(p)$ are the form factors  \eqref{a1} and \eqref{a2} defined in the momentum space.
For the second determinant on the right-hand side of Eq. \eqref{det}, the result is not as clear as the first. As the first step, we shall expand it into a power series, i.e.,
\begin{eqnarray}
    \nonumber\det\left[\hat{I}+\frac{p^2 a_2(p)(i\slashed{p})}{p^2 a_1(p)+m^2}\right]&=&e^{{\rm Tr}\ln{\left[\hat{I}+\frac{p^2 a_2(p)(i\slashed{p})}{p^2 a_1(p)+m^2}\right]}}\\
    &=&e^{{\rm Tr}\sum\limits_{n=1}^{\infty}\frac{(-1)^{n+1}}{n}\left[\frac{p^2 a_2(p)(i\slashed{p})}{p^2 a_1(p)+m^2}\right]^n}.
\end{eqnarray}
Observe that any combination of an odd number of gamma matrices leads to zero traces, so only their even combinations yield a non-trivial trace. Hence, we have
\begin{eqnarray}
    \nonumber\det\left[\hat{I}+\frac{p^2 a_2(p)(i\slashed{p})}{p^2 a_1(p)+m^2}\right]&=&e^{-\frac{1}{2}{\rm Tr}\left\{\sum\limits_{k=1}^{\infty}\frac{1}{k}\left[\left(\frac{p^2 a_2(p)}{p^2 a_1(p)+m^2}\right)^2 p^2\right]^{k}\hat{I}\right\}}\\
    &=&e^{\ln{\left[1-\left(\frac{p^3 a_2(p)}{p^2 a_1(p)+m^2}\right)^2\right]^2}}.
\end{eqnarray}
Thus,
\begin{equation}
    \det\left[\hat{I}+\frac{p^2 a_2(p)(i\slashed{p})}{p^2 a_1(p)+m^2}\right]=\left[1-\left(\frac{p^3 a_2(p)}{p^2 a_1(p)+m^2}\right)^2 \right]^2.
    \label{det2}
\end{equation}
Substituting Eqs. \eqref{det1} and \eqref{det2} into Eq. \eqref{det}, we arrive at the following dispersion relation:
\begin{eqnarray}
p^2a_1(p)+m^2=\pm \,p^3 a_2(p).
\end{eqnarray}

\begin{figure}[h!]
    \centering
    \includegraphics[width=0.7\linewidth]{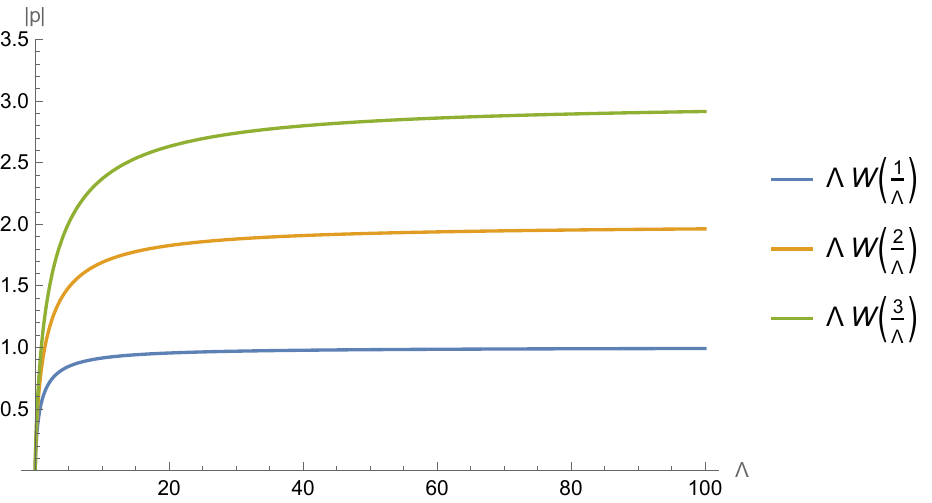}
    \caption{This figure shows different plots of \eqref{Lam} for different mass values: $m=1$ (blue line), $m=2$ (yellow line) and $m=3$ (green line).}
    \label{fig:enter-label}
\end{figure}

The aforementioned dispersion relation is obviously non-polynomial since $a_1(p)$ and $a_2(p)$ are nonlocal coefficients defined in the momentum space. As a result, in general we cannot guarantee
analytic solutions, although they exist for special cases. To illustrate them, we consider the nonlocal coefficients given by \eqref{a3} and \eqref{a4}. In this case, the dispersion relation reduces to
\begin{eqnarray}
   |p|^2 e^{\pm 2|p|/\Lambda}=m^2, 
\end{eqnarray}
where $|p|^2=(E^2-|\vec{p}|^2)$. This equation recovers the standard (local) dispersion relation when $\Lambda\to\infty$, as expected. Furthermore, it presents analytical solutions given by
\begin{equation}
    |p|=\Lambda\, W_{0}\left(\frac{m}{\Lambda}\right),
    \label{Lam}
\end{equation}
where $W_{0}\left(z\right)$ is the principal branch of the Lambert W function (also called product logarithm), $W(z)$, which is a multivalued complex function, defined as the inverse of $f(z)=z\,e^{z}$, with $z\in \mathbb{C}$. These functions have an infinite number of branches $W_k (z)$, where $k\in \mathbb{Z}$ \cite{Mathematica}. However, since we are interested only in real solutions, then there could be two possible branches, namely: $W_{0}(z)$ defined in the range $z\geq -e^{-1}$ and $W_{-1}(z)$ defined in the range $0>z\geq-e^{-1}$. Nonetheless, we restrict our solution to the principal branch because the lower branch, $W_{-1}(z)$, is not well-defined at $z=0$.  Therefore, it ensures that $|p|$ and $m^2$ are positive, as it must be. In Fig.\eqref{fig:enter-label}, we plot the behavior of the solution \eqref{Lam}. It is worth noting that as $\Lambda$ grows, $|p|$ approaches $m$, recovering the standard (local) result, as expected. To shed more light on this, let us expand the dispersion relation \eqref{Lam} around $\Lambda\to\infty$, yielding
 \begin{equation}
     E^2=|\vec {p}|^2+m^2- \frac{2m^3}{\Lambda}+\mathcal{O}\left(\frac{1}{\Lambda^2}\right).
     \label{DRmo}
 \end{equation}
 This equation displays the first-order correction due to the nonlocal effects. Furthermore, it remains invariant under parity transformations: $E\to -E$ and $|\vec{p}|\to -|\vec{p}|$, differently from Lorentz-breaking models which their dispersion relations break spontaneously parity transformations \cite{Colladay:1996iz,Kostelecky:2000mm,Carone}. Additionally, our results are in accordance with quantum-gravity phenomenology models, in which the standard dispersion relation is modified by including Planck-suppressed terms \cite{Amelino1,Alfaro,Magueijo} similar to that in Eq. \eqref{DRmo}.  

\section{One-loop nonlocal fermionic effective action with Yukawa coupling}

In this section, we aim to compute the one-loop nonlocal fermionic effective action in the presence of a Yukawa coupling. To begin with, we add the Yukawa coupling to the nonlocal spinor Lagrangian \eqref{eq1} to yield

\begin{eqnarray}   
\mathcal{L}_{Y}=\bar\Psi[i\slashed{\partial} f(\slashed{\partial})-m+\beta\Phi]\Psi, 
    \label{yukawa}
\end{eqnarray}
where $\beta$ is the Yukawa coupling constant and $\Phi$ is the scalar field. The field equation associated with the Yukawa Lagrangian is
\begin{equation}
[i\slashed{\partial} f(\slashed{\partial})-m+\beta\Phi]\Psi=0,    
\end{equation}
which can be rewritten as
\begin{equation}
    [\slashed{\partial}^{2} f^{2}(\slashed{\partial})+(m-\beta\Phi)^2]\Psi=0.
\end{equation}

In the following, we turn our attention to the computation of the fermionic one-loop effective action.

\subsection{Fermionic effective action}

Let us briefly review the background field method, a nice approach to dealing with quantum corrections around the classical field. To do that, the first ingredient is to allow the dynamical fields to fluctuate around the background fields, we mean, $\Psi\to \Psi+\sqrt{\hslash}\psi$ and $\bar{\Psi}\to\bar{\Psi}+\sqrt{\hslash}\bar{\psi}$, where $\Psi$ and $\bar{\Psi}$ are the background fields, while $\psi$ and $\bar{\psi}$ are the quantum fields. The second ingredient is the fermionic effective action $\Gamma[\Psi,\bar{\Psi}]$, defined as the generating functional of one-particle-irreducible Green functions (1PI) 
\begin{equation}
    e^{i\frac{\Gamma[\Psi, \bar{\Psi}]}{\hslash}}=\int \mathcal{D}\bar{\psi}\mathcal{D}\psi\,e^{\frac{i}{\hslash}(S[\Psi+\sqrt{\hslash}\psi,\bar{\Psi}+\sqrt{\hslash}\bar{\psi}]+(\bar{\eta}\psi+\bar{\psi}\eta))},
\label{GE}
\end{equation}
where $\bar{\eta}$ and $\eta$ are Grassmann sources for the quantum fields, $\psi$ and $\bar{\psi}$, respectively. Integrating out overall quantum field configurations is not a simple task, so it is more convenient to expand the effective action in power series of $\hslash$ such that
\begin{equation}
    \Gamma[\Psi, \bar{\Psi}]=S[\Psi, \bar{\Psi}]+\hslash\,\Gamma^{(1)}[\Psi, \bar{\Psi}]+\mathcal{O}(\hslash^2).
\end{equation}
The first-order term  $\Gamma^{(1)}[\Psi, \bar{\Psi}]$ is the one-loop fermionic effective action whose explicit form reads
\begin{eqnarray}
    \nonumber\Gamma^{(1)}[\Psi, \bar{\Psi}]&=&-\frac{i}{2} \ln\int \mathcal{D}\bar{\psi}\mathcal{D}\psi e^{i\bar{\psi}\Delta\psi}\\\
   \nonumber &=&-\frac{i}{2}\ln \det \Delta\\
   &=&-\frac{i}{2}{\rm Tr}\ln\Delta
\end{eqnarray}
where $\Delta=\dfrac{\delta S}{\delta \bar\psi \delta\psi}\Bigg|_{\bar{\psi}=\bar{\Psi},\psi=\Psi}$ is the kernel of the quadratic operator of the classical action.

We are now able to compute the one-loop fermionic effective action. In addition, let us assume that the spinor background fields vanish because we are interested in capturing the nonlocal effects due to their quantum fluctuations only. Now, using the background field method to the nonlocal action \eqref{yukawa}, the one-loop contribution to the fermionic effective action is given by 
\begin{eqnarray}
\Gamma^{(1)}&=&-i{\rm Tr}\ln[i\slashed{\partial} f(\slashed{\partial})-m+\beta\Phi]=
-\frac{i}{2}{\rm Tr}\ln\left([i\slashed{\partial} f(\slashed{\partial})-\tilde{\Phi}][-i\slashed{\partial} f(\slashed{\partial})-\tilde{\Phi}]\right)
=\nonumber\\
&=&-\frac{i}{2}{\rm Tr}\ln\left(\slashed{\partial}^2f^2(\slashed{\partial})+\tilde{\Phi}^2\right),
\end{eqnarray}
where we have made the field redefinition $\tilde{\Phi}=m-\beta\Phi$.

Using the results of the previous section,  we can write
\begin{eqnarray}
\Gamma^{(1)}&=&-\frac{i}{2}{\rm Tr}\ln\left( a^{\prime}_1(\Box)+a^{\prime}_2(\Box)\gamma^{\mu}\partial_{\mu}+\tilde{\Phi}^2\right)=\nonumber\\
&=&-\frac{i}{2}{\rm Tr}\ln\left( a^{\prime}_1(\Box)+\tilde{\Phi}^2\right)-\frac{i}{2}{\rm Tr}
\ln\left(1+\frac{a^{\prime}_2(\Box)\gamma^{\mu}\partial_{\mu}}{a^{\prime}_1(\Box)+\tilde{\Phi}^2}\right),
\label{G}
\end{eqnarray}
where,
\begin{eqnarray}
    a^{\prime}_1(\Box)&=&-\Box\, a_1(\Box);\\   a^{\prime}_2(\Box)&=&-\Box\,a_{2}(\Box).
\end{eqnarray}
In order to have explicit results, one must explicitly have the nonlocal operator forms. In particular, we use those given by Eqs. \eqref{a3} and \eqref{a4}. In this case,  one has
\begin{eqnarray}
\label{ap1} a^{\prime}_1(\Box)&=&-\Box\cosh{\left(\frac{2\sqrt{-\Box}}{\Lambda}\right)};\\
\label{ap2}a^{\prime}_2(\Box)&=&-\sqrt{-\Box}\sinh{\left(\frac{2\sqrt{-\Box}}{\Lambda}\right)}.
\end{eqnarray}

\subsection{On Wick rotation versus Euclidean formulation in nonlocal theories}
\label{subsec:wick}

This subsection is aimed at clarifying an important formal issue that distinguishes nonlocal theories from local ones.
In local quantum field theory, the term Wick rotation is commonly used in two equivalent ways:

\begin{enumerate}
\item[(A)] \emph{Contour deformation in the complex momentum plane,}
e.g., $p^0 \to i p^0_E$, justified by analyticity and the possibility of rotating the
integration contour without crossing singularities.

\item[(B)] \emph{Euclidean formulation,} defined by $t=-i\tau$ (or,
equivalently, by working directly in Euclidean signature), so that the Minkowski metric
is replaced by a Euclidean metric and one computes correlators from Euclidean
loop integrals.
\end{enumerate}

For local theories, (A) and (B) lead to equivalent results under standard assumptions.
For nonlocal theories, instead, they need not be equivalent in general. The reason is that
entire form factors typical of ghost-free nonlocal models can spoil the analyticity
properties required to deform integration contours in the complex momentum plane, in other words, they may induce essential
singularities at infinity and/or uncontrolled exponential growth along certain complex directions. In such
cases, the contour rotation (A) is not guaranteed to be well-defined because the contribution from arcs at infinity does not necessarily vanish, and the integration boundaries cannot be smoothly deformed. As a simple example, consider the form factor $h(\Box)=\Box/\Lambda^2$ in Eq.\eqref{EQ2new}, thus the associated two-point Green function (propagator) of this model is simply given by 
\begin{equation}
    \langle\Phi(-p)\Phi(p)\rangle=-i\frac{e^{+p^2/\Lambda^2}}{p^2},
\end{equation}
and an associated typical loop integral is of the form
\begin{equation}
    I=\int_{-\infty}^{+\infty} dp^0\, \frac{e^{(-p_{0}^2+\vec{p}^2)/\Lambda^2}e^{-ip_0x_0+i\vec{p}\cdot\vec{x}}}{p^2},
\end{equation}
which diverges along certain directions in the complex plane $p^{0}$. Note that the divergences arise whenever $|\mbox{Im}|(p_0)>|\mbox{Re}(p_0)|$.

This subtlety is relevant for nonlocal models. In particular, as already noted below
Eq.~(52), the standard Wick rotation used to evaluate loop integrals in Minkowski spacetime can fail due to divergences of the form factors along certain directions in the
complex $p^0$ plane, which compromises the usual analyticity arguments and may render correlators ill-defined in Mikowski spacetime \cite{Buoninfante:2019zws}.
Therefore, in this work we do not interpret the change to the Euclidean space as a
literal contour rotation in the complex plane. Instead, we adopt procedure (B): we define
the Euclidean theory via imaginary time (equivalently, Euclidean loop integrals),
compute the one-loop effective action in Euclidean space, and only at the end (when it is
well-defined) consider analytic continuation back to real time for the specific observables
of interest.

Concretely, within the computation of quantum corrections,  we work in Euclidean space with
$\partial_\mu=(\partial_{x_4},\vec{\nabla})$ and $\Box = \partial_{x_4}^2+\nabla^2$, where $x^4$ is a space-like dimension, and use
Euclidean gamma matrices satisfying $\{\gamma_\mu,\gamma_\nu\}=-2\delta_{\mu\nu}$.
Within this Euclidean formulation, the traces in Eq.~(52) are well-defined and lead to
the Euclidean one-loop effective action which is recovered using the prescription
$i\Gamma^{(1)}\rightarrow\tilde{\Gamma}^{(1)}$, where the tilde indicates the  Euclidean physical quantities. Thus, the Euclidean one-loop effective action is explicitly given by
\begin{equation}
    \tilde{\Gamma}^{(1)}=\frac{1}{2}{\rm Tr}\ln\left( a^{\prime}_1(\Box)+\tilde{\Phi}^2\right)+\frac{1}{2}{\rm Tr}
\ln\left(1-i\frac{a^{\prime}_2(\Box)\gamma^{\mu}\partial_{\mu}}{a^{\prime}_1(\Box)+\tilde{\Phi}^2}\right).\label{klj}
\end{equation}

The first term of Eq.\eqref{klj}, which we denote as $\tilde\Gamma^{(1)}_1$, so that it involves only $a^{\prime}_1(\Box)$, can be evaluated directly through Fourier transform, thereby,
\begin{eqnarray}
   \nonumber  \tilde\Gamma^{(1)}_1&=&\frac{1}{2}\int\frac{d^4p}{(2\pi)^4}\ln\left[p^2\cosh\left(\frac{2p}{\Lambda}\right)+\tilde{\Phi}^2\right]\\
   &=&\frac{1}{16\pi^2}\int dp\,p^3\ln\left[p^2\cosh\left(\frac{2p}{\Lambda}\right)+\tilde{\Phi}^2\right].
\end{eqnarray}
This integral is very cumbersome to evaluate exactly. Therefore, we
perform the following approximation (similar to the calculation scheme adopted in \cite{Buchbinder:2001hn}), namely,
\begin{eqnarray}  \tilde\Gamma^{(1)}_1=I_{IR}^{(1)}+I_{UV}^{(1)},
\end{eqnarray}
where 
\begin{eqnarray}
 \nonumber   I_{IR}^{(1)}&=&\frac{1}{16\pi^2}\int_{0}^{\Omega}dp\,p^3\left(\ln(p^2+\tilde{\Phi}^2)+\frac{2p^4}{(\tilde{\Phi}^2+p^2)\Lambda^2}\right)+\mathcal{O}\left(\frac{1}{\Lambda^4}\right)\\
  \nonumber  &=&\frac{1}{16\pi^2}\Biggl[\frac{1}{8} \Biggl(4 \tilde{\Phi}^4 \ln (\tilde{\Phi})+2 \tilde{\Phi}^2 \Omega ^2+2 \left(\Omega ^4-\tilde{\Phi}^4\right) \ln \left(\tilde{\Phi}^2+\Omega ^2\right)\Biggr)+\\
  \nonumber &+&\frac{6 \tilde{\Phi}^4
   \Omega ^2-3 \tilde{\Phi}^2 \Omega ^4+6 \tilde{\Phi}^6 \ln \left(\frac{\tilde{\Phi}^2}{\tilde{\Phi}^2+\Omega ^2}\right)}{6 \Lambda^2}
    +\frac{12 \tilde{\Phi}^6 \ln \left(\frac{\tilde{\Phi}^2+\Omega ^2}{\tilde{\Phi}^2}\right)-\frac{12 \tilde{\Phi}^6 \Omega ^2+6 \tilde{\Phi}^4 \Omega ^4-2 \tilde{\Phi}^2 \Omega ^6+\Omega ^8}{\tilde{\Phi}^2+\Omega ^2}}{ 3
 \Lambda^2}\Biggr]+\\
   &+&\mathcal{O}\left(\frac{1}{\Lambda^4}\right),
    \label{ir1}
\end{eqnarray}
with $\Omega$ being an arbitrary intermediate-mass scale satisfying $\tilde{\Phi}<\Omega<\Lambda$. This integral is carried out over small momenta (IR limit), which is a good approximation to the exact result when $p\sim \tilde{\Phi}<\Omega\ll \Lambda$ holds, and
\begin{eqnarray}
    \nonumber I^{(1)}_{UV}&=&\frac{\tilde{\Phi}^2}{8\pi^2}\int_{\Omega}^{\infty}dp\, p\,e^{\frac{-2p}{\Lambda}}+\,\ ...\\
    &=&\frac{\tilde{\Phi}^2 }{32\pi^2}\left(e^{-\frac{2 \Omega }{\Lambda}} (\Lambda^2+2 \Omega\Lambda )\right) +\,\ ....
    \label{uv1}
\end{eqnarray}
The ellipses mean irrelevant or subleading terms.
 It should be noted that this integral is a good approximation to the exact result as $\tilde{\Phi}\ll\Lambda<p$. Combining both Eqs. \eqref{ir1} and \eqref{uv1}, one finds
 \begin{eqnarray}
    \nonumber \tilde\Gamma^{(1)}_1 &=&\frac{1}{16\pi^2}\Biggl[\frac{1}{8} \Biggl(4 \tilde{\Phi}^4 \ln (\tilde{\Phi})+2 \tilde{\Phi}^2 \Omega ^2+2 \left(\Omega ^4-\tilde{\Phi}^4\right) \ln \left(\tilde{\Phi}^2+\Omega ^2\right)\Biggr)+\\
  \nonumber&+&\frac{6 \tilde{\Phi}^4
   \Omega ^2-3 \tilde{\Phi}^2 \Omega ^4+6 \tilde{\Phi}^6 \ln \left(\frac{\tilde{\Phi}^2}{\tilde{\Phi}^2+\Omega ^2}\right)}{6 \Lambda^2}
    \Biggr]+\frac{12 \tilde{\Phi}^6 \ln \left(\frac{\tilde{\Phi}^2+\Omega ^2}{\tilde{\Phi}^2}\right)-\frac{12 \tilde{\Phi}^6 \Omega ^2+6 \tilde{\Phi}^4 \Omega ^4-2 \tilde{\Phi}^2 \Omega ^6+\Omega ^8}{\tilde{\Phi}^2+\Omega ^2}}{ 3
 \Lambda^2}\\
    &+&\frac{\tilde{\Phi}^2 }{32\pi^2}\left(e^{-\frac{2 \Omega }{\Lambda}} (\Lambda^2+2 \Omega\Lambda )\right).\label{G1}
 \end{eqnarray} 

As for the second term of Eq.\eqref{klj}, we proceed with it now. Expanding it in power series, we have
\begin{eqnarray}
\tilde\Gamma^{(1)}_2&=&\frac{1}{2}{\rm Tr}\ln\left(1-i\frac{a^{\prime}_2(\Box)\gamma^{\mu}\partial_{\mu}}{a^{\prime}_1(\Box)+\tilde{\Phi}^2}\right)=
\nonumber\\
&=&\frac{1}{2}{\rm Tr}\sum\limits_{n=1}^{\infty}\frac{(-1)^{n+1}}{n}\left[-i\frac{a^{\prime}_2(\Box)\gamma^{\mu}\partial_{\mu}}{a^{\prime}_1(\Box)+\tilde{\Phi}^2}\right]^n.
\end{eqnarray}
Then, we take into account that only even degrees of $\gamma^{\mu}$ yield a non-zero trace, we write
\begin{eqnarray}
\tilde\Gamma^{(1)}_2=-\frac{1}{2}{\rm Tr}\sum\limits_{k=1}^{\infty}\frac{1}{2k}\left[\frac{a^{\prime}_2(\Box)}{a^{\prime}_1(\Box)+\tilde{\Phi}^2}\right]^{2k}\Box^k=-\frac{1}{2}{\rm Tr}\sum\limits_{k=1}^{\infty}\frac{1}{2k}\left[\left[\frac{a^{\prime}_2(\Box)}{a^{\prime}_1(\Box)+\tilde{\Phi}^2}\right]^2\Box\right]^k.
\end{eqnarray}
Here we used the fact that $(\gamma^{\mu}\partial_{\mu})^2=-\Box$.
This series is easily summed up as
\begin{eqnarray}
\tilde\Gamma^{(1)}_2&=&-\frac{1}{2}{\rm Tr}\sum\limits_{k=1}^{\infty}\frac{1}{2k}\left[\left[\frac{a^{\prime}_2(\Box)}{a^{\prime}_1(\Box)+\tilde{\Phi}^2}\right]^2\Box\right]^k=\frac{1}{4}\ln\left[1+\left[ \frac{a^{\prime}_2(\Box)}{a^{\prime}_1(\Box)+\tilde{\Phi}^2}\right]^2\Box\right].
\end{eqnarray}
 With this at hand, the next step is to perform the Fourier transform. By doing so,  one arrives at
\begin{eqnarray}
    \tilde\Gamma^{(1)}_2=\frac{1}{4}\int \frac{d^4p}{(2\pi)^4}\ln\left[1-\left[ \frac{p\,a^{\prime}_2(p^2)}{a^{\prime}_1(p^2)+\tilde{\Phi}^2}\right]^2 \right].
    \label{effc}
\end{eqnarray}
Now, plugging Eqs. \eqref{ap1} and \eqref{ap2} into \eqref{effc}, one finds
\begin{eqnarray}
    \nonumber \tilde\Gamma^{(1)}_2&=&\frac{1}{4}\int \frac{d^4p}{(2\pi)^4}\ln\left[1-\left[ \frac{p^2\sinh\left(\frac{2p}{\Lambda}\right)}{p^2\cosh\left(\frac{2p}{\Lambda}\right)+\tilde{\Phi}^2}\right]^2 \right]\\
    &=&\frac{1}{32\pi^2}\int_{0}^{\infty} dp\,p^3\ln\left[1-\left[ \frac{p^2\sinh\left(\frac{2p}{\Lambda}\right)}{p^2\cosh\left(\frac{2p}{\Lambda}\right)+\tilde{\Phi}^2}\right]^2 \right].
\end{eqnarray}
This integral evidently diverges in the UV limit. So, we need to regularize it, which means that a regularization scheme should be adopted. We use a UV cutoff regularization scheme by defining the UV regulator, $\Lambda_{UV}$, fulfilling the following feature $\tilde{\Phi}\ll\Omega\ll\Lambda\leq\Lambda_{UV}$. Note that the previous integral cannot be analytically solved so that we perform a similar approximation to $\tilde\Gamma_1^{(1)}$, that is,
\begin{eqnarray}
\tilde\Gamma_{2}^{(1)}=I^{(2)}_{IR}+I^{(2)}_{UV},
\end{eqnarray}
where 
\begin{eqnarray}
   \nonumber I^{(2)}_{IR}&=&-\frac{1}{8\pi^2\Lambda^2}\int_{0}^{\Omega}dp\frac{p^9}{(p^2+\tilde{\Phi}^2)}+\mathcal{O}\left(\frac{1}{\Lambda^4}\right)\\
   \nonumber &=&-\frac{1}{8 \pi ^2 \Lambda^2} \left[2\tilde{\Phi}^6 \ln \left(\frac{\tilde{\Phi}^2}{\tilde{\Phi}^2+\Omega^2}\right)+\frac{12 \tilde{\Phi}^6 \Omega^2+6 \tilde{\Phi}^4 \Omega^4-2 \tilde{\Phi}^2 \Omega^6+\Omega^8}{6 \left(\tilde{\Phi}^2+\Omega^2\right)}\right]+\\&+&\mathcal{O}\left(\frac{1}{\Lambda^4}\right)
   \label{ir}
\end{eqnarray}
corresponds to the integration over small momenta (IR limit), which is a good approximation to the exact result when $p\leq \tilde{\Phi}\sim \Omega\ll \Lambda$ holds, and
\begin{eqnarray}
    \nonumber I^{(2)}_{UV}&=&\frac{\tilde{\Phi}^2}{32\pi^2}\int_{\Omega}^{\Lambda_{UV}}dp\, p\,e^{\frac{2p}{\Lambda}}+...\\
    &=&\frac{\tilde{\Phi}^2\Lambda }{128\pi^2}\left(-e^{\frac{2 \Lambda_{UV} }{\Lambda}} (\Lambda-2\Lambda_{UV} )+e^{\frac{2\Omega}{\Lambda}}(\Lambda-2\Omega)\right) +...
    \label{uv2}
\end{eqnarray}
with $\Lambda_{UV}$ being an UV cutoff regulator mass scale. The integral $I^{(2)}_{UV}$ is performed over high momenta (UV limit) and displays a divergent term that can be renormalized by adding a counterterm in the action. 
  It should be noted that this integral is a good approximation to the exact result as $\tilde{\Phi}\ll\Omega<\Lambda\leq p \sim \Lambda_{UV}$. Combining both Eqs. \eqref{ir} and \eqref{uv2}, one finds
  \begin{eqnarray}
     \nonumber\tilde\Gamma_{2}^{(1)}&=& \frac{1}{8 \pi ^2 \Lambda^2} \left[2\tilde{\Phi}^6 \ln \left(\frac{\tilde{\Phi}^2}{\tilde{\Phi}^2+\Omega^2}\right)+\frac{12 \tilde{\Phi}^6 \Omega^2+6 \tilde{\Phi}^4 \Omega^4-2 \tilde{\Phi}^2 \Omega^6+\Omega^8}{6 \left(\tilde{\Phi}^2+\Omega^2\right)}\right]+\\
     &+&\frac{\tilde{\Phi}^2\Lambda }{128\pi^2}\left(-e^{\frac{2 \Lambda_{UV} }{\Lambda}} (\Lambda-2\Lambda_{UV} )+e^{\frac{2\Omega}{\Lambda}}(\Lambda-2\Omega)\right).
     \label{G2}
  \end{eqnarray}
Therefore, the full effective action \eqref{G} is simply given by the sum of Eqs. \eqref{G1} and \eqref{G2}. 

Let us investigate the IR and UV limits for more insight into the physical results. Observe that in the IR regime, i.e., when $\Omega\gg\tilde{\Phi}$, the fermionic one-loop effective action reduces to 
\begin{eqnarray}
    \nonumber \tilde\Gamma_{IR}^{(1)}&=& I_{IR}^{(1)}+I_{IR}^{(2)}\\
    &=&
    \frac{\tilde{\Phi}^4 \ln \left(\frac{\tilde{\Phi}}{\Omega}\right)}{32 \pi ^2}+\frac{\tilde{\Phi}^2 \Omega ^2}{32\pi^2}+\frac{\Omega^2}{8\pi^2\Lambda^2}\left(\frac{\tilde{\Phi}^2 \Omega ^2}{4}-\tilde{\Phi}^4\right)- \frac{3\tilde{\Phi}^6 \ln \left(\frac{\tilde{\Phi}}{\Omega}\right)}{8 \pi ^2 \Lambda^2} 
 +\mathcal{O}\left(\frac{1}{\Lambda^4}\right).
\end{eqnarray}
It is important to remark that in the IR limit the one-loop effective action is highly suppressed by the nonlocal mass scale ($\Lambda\to\infty$). Accordingly, in this case, the leading contributions to the one-loop effective action come from the local term, as expected.

On the other hand, to probe the UV regime, it is reasonable to take the UV regulator to be of the order of the nonlocality mass scale, $\Lambda_{UV}=\Lambda$. In this scenario, the one-loop effective action reads
\begin{eqnarray}
\nonumber\tilde\Gamma_{UV}^{(1)}&=&I_{UV}^{(1)}+I_{UV}^{(2)}\\
    &=&
    \frac{(5+e^2)\Lambda^2}{128\pi^2}\tilde{\Phi}^2+\mbox{subleading terms}.
\end{eqnarray}
This means that the nonlocal contributions to the fermionic one-loop effective action become very relevant in the UV limit.

\section{Comments on coupling of nonlocal spinor within the electromagnetic field}

This section aims to construct a gauge invariance nonlocal spin-$1/2$ theory minimally coupled to an electromagnetic field. To do that, we follow the traditional recipe employed in quantum field theory, which is to promote the partial derivative to the covariant one through the following prescription, $\partial_{\mu}\rightarrow D_{\mu}=\partial_{\mu}+ie A_{\mu}$, where  $A_{\mu}$ is the $U(1)$ gauge field.

In this case, based on the previous section, the form factor takes the form $f(\slashed{D})=e^{-\frac{\slashed{D}}{\Lambda}}$, where $\slashed{D}\equiv \gamma^{\mu}D_{\mu}$ is the Dirac operator. Therefore, the nonlocal interacting spin-$1/2$ action in Minkowski spacetime reads 
\begin{equation}
       \mathcal{S}_{I}=\int d^{4}x\,\bar\Psi \left(i\slashed{D} f(\slashed{D})-m\right)\Psi.
    \label{gauge1}
\end{equation}
It is well-known that the version of the above action is invariant under $U(1)$ gauge transformations,
\begin{equation}
\begin{aligned}
       &A_{\mu} \xrightarrow{U(1)}A^{'}_{\mu}=A_{\mu}+\partial_{\mu}\alpha(x);\\
      & \Psi \xrightarrow{U(1)} \Psi^{'}= e^{-ie\alpha(x)}\Psi;\\
       &\bar{\Psi} \xrightarrow{U(1)} \bar{\Psi}^{'}=\bar{\Psi} e^{ie\alpha(x)}.
       \label{gauge}
\end{aligned}
\end{equation}
However, it is not obvious that the nonlocal action \eqref{gauge1} is, since the form factor is not a linear function of the Dirac operator. In fact, it is natural to expect some tensions when combining local symmetries with nonlocal theories. To check this out, one must show that $i\slashed{D}f(\slashed{D})\Psi$ transforms covariantly under \eqref{gauge}, that is, $i\slashed{D}f(\slashed{D})\Psi \xrightarrow{U(1)}
[i\slashed{D}f(\slashed{D})\Psi]^{'}=e^{-iq\alpha(x)}i\slashed{D}f(\slashed{D})\Psi$. After a simple algebraic manipulation, we see that
\begin{eqnarray}
  \left (i\slashed{D}f(\slashed{D})\Psi\right)^{'} =\left[i\gamma^{\mu}D_{\mu}\left(1-\frac{\gamma^{\nu_{1}}D_{\nu_{1}}}\Lambda+\frac{\gamma^{\nu_{1}}\gamma^{\nu_{2}}D_{\nu_{1}}D_{\nu_{2}}}{2\Lambda^{2}}+\cdots\right)\Psi\right]^{'}.
  \label{35}
\end{eqnarray}
The zeroth-order term in the above power series is just the local case, which transforms covariantly under $U(1)$ gauge transformations, i.e,
\begin{equation}
\begin{aligned}
 \left(i\gamma^{\mu}D_{\mu}\Psi\right)^{'} =
   e^{-i q \alpha(x)}i \gamma^{\mu}D_{\mu}\Psi.
\end{aligned}
\end{equation}
The first-order term transforms as:
\begin{equation}
\begin{aligned}
    \left[i\gamma^{\mu}D_{\mu}\left(-\frac{\gamma^{\nu_{1}}D_{\nu_{1}}}\Lambda\right)\Psi\right]^{'} &=-\frac{i\gamma^{\mu}\gamma^{\nu_{1}}}{\Lambda}(D_{\mu}D_{\nu_{1}}\Psi)^{'}\\ &=-\frac{i\gamma^{\mu}\gamma^{\nu_{1}}}{\Lambda}(\partial_{\mu}+i q A_{\mu}+i q \partial_{\mu}\alpha(x)) e^{-iq \alpha(x)}D_{\nu_{1}}\Psi\\ &= -\frac{i\gamma^{\mu}\gamma^{\nu_{1}}}{\Lambda}e^{-iq\alpha(x)}(\cancel{-i q \partial_{\mu}\alpha(x)}+\partial_{\mu}+i q A_{\mu}+\cancel{i q \partial_{\mu}\alpha(x)})D_{\nu_{1}}\Psi\\ &= e^{-iq\alpha(x)}i\gamma^{\mu}D_{\mu}\left(-\frac{\gamma^{\nu_{1}}D_{\nu_{1}}}{\Lambda}\right)\Psi.
\end{aligned}   
\end{equation}
Generically, the $j$-th term transforms as:
\begin{equation}
\begin{aligned}
\left[i\gamma^{\mu}D_{\mu}\left(\frac{(-1)^{j}(\gamma^{\nu}D_{\nu})^{j}}{j!\Lambda^{j}}\Psi\right)\right]^{'}&=\frac{(-1)^{j}}{j!\Lambda^{j}}i\gamma^{\mu}\gamma^{\nu_{1}}\ldots\gamma^{\nu_{j}}(D_{\mu}D_{\nu_{1}}\ldots D_{\nu_{j}} \Psi)^{'}\\&= \frac{(-1)^{j}}{j!\Lambda^{j}}i\gamma^{\mu}\gamma^{\nu_{1}}\ldots\gamma^{\nu_{j}}D_{\mu}^{'}D_{\nu_{1}}^{'}\ldots e^{-iq\alpha(x)} \left(\partial_{\nu_{j-1}}+iq A_{\nu_{j-1}}\right) D_{\nu_{n}}\Psi\\&
.\\
&.\\
&.\\
&=e^{-iq\alpha(x)}i\gamma^{\mu}D_{\mu}\left(\frac{(-1)^{j}(\gamma^{\nu}D_{\nu})^{j}}{j!\Lambda^{j}}\right)\Psi.
\end{aligned}   
\label{last}
\end{equation}
Then, plugging Eq. \eqref{last} into Eq. \eqref{35}, one finds that $i\slashed{D}f(\slashed{D})\Psi$ transforms covariantly under $U(1)$ gauge transformations.

The action \eqref{gauge1} yields the following field equation 
\begin{equation}
        [i\slashed{D} f(\slashed{D})-m]\Psi=0.
    \label{equ2}
\end{equation}
Proceeding with the trick of ``squaring'' the previous equation, we obtain  
\begin{equation}
        [\slashed{D}^{2} f^{2}(\slashed{D})+m^{2}]\Psi=0
    \label{kleinb}.
\end{equation}
  Using the gamma matrices properties, one can rewrite
\begin{equation}
\begin{aligned}
        \slashed{D}^{2} &=-D^{2}\hat{I}+\frac{e }{2} \sigma^{\mu\nu}F_{\mu\nu}\\
        &=-\left[\Box+ ie (\partial_{\mu}A^{\mu} + A_{\mu}\partial^{\mu})- e^{2}A_{\mu}A^{\mu}\right]\hat{I}+\frac{e }{2} \sigma^{\mu\nu}F_{\mu\nu},   
\end{aligned}
    \label{b2}
\end{equation}
where $F_{\mu\nu}=\partial_{\mu}A_{\nu}-\partial_{\nu}A_{\mu}$ is the field strength and $\sigma^{\mu\nu}=\frac{i}{2}[\gamma^{\mu},\gamma^{\nu}]$. Therefore, the field equation can be compactly written as follows
\begin{equation}
    \left[\left(D^{2}\hat{I}-\frac{e \sigma^{\mu\nu}}{2} F_{\mu\nu}\right) f^{2}(\slashed{D})  -m^{2}\right]\Psi=0.
    \label{interac}
\end{equation}

The same arguments to the free case are applied here concerning the form factor; thus, it should be an entire function represented as a series of powers of $\slashed{D}$, i.e., 
\begin{equation}
        f(\slashed{D})=\sum
        \limits_{n=0}^{\infty} b_{n} \slashed{D}^{n}_{\Lambda},
    \label{fb}
\end{equation}
where $\slashed{D}_{\Lambda}\equiv \dfrac{\slashed{D}}{\Lambda}$ is dimensionless and $b_{n}$ are the coefficients (dimensionless) of the power series, with initial condition $b_0=1$. Similar to Section \eqref{sec2}, let us consider two kinds of form factors:
\begin{eqnarray}
    f_{I}(\slashed{D})&=&e^{-\frac{\slashed{D}}{\Lambda}};\\
    \label{II}f_{II}(\slashed{D})&=&e^{-i\frac{\slashed{D}}{\Lambda}}.
\end{eqnarray}
However, note that, unlike the free case, the spectrum of the square of the interacting Dirac operator is more involved. Alternatively, we write the nonlocal interacting Dirac equation in terms of the spectrum of the operator $\slashed{\partial}$. Hence, using the Baker-Campbell-Hausdorff (BCH) formula, Eq. \eqref{interac} takes the form,  
\begin{eqnarray}
    \nonumber&\Biggl\{&\left(D^{2}\hat{I}-\frac{e \sigma^{\mu\nu}}{2} F_{\mu\nu}\right) \bigg[e^{-\frac{2ie\slashed{A}}{\Lambda}}\left(\cosh{\left(\frac{2\lambda}{\Lambda}\right)}\hat{I}-\frac{1}{\lambda}\sinh{\left(\frac{2\lambda}{\Lambda}\right)}\slashed{\partial}\right)-\\
    &-&e^{\frac{2ie}{\Lambda^2}\left(-2i\sigma^{\mu\nu}A_\mu \partial_\nu-\frac{i}{2}\sigma^{\mu\nu}F_{\mu\nu}+\partial_{\mu}A^{\mu}\right)}+ \mathcal{O}\left(e^\frac{e^2}{\Lambda^3}\right)\bigg]  -m^{2}\Biggr\}\Psi=0,
\end{eqnarray}
where $\mathcal{O}\left(e^\frac{e^2}{\Lambda^3}\right)$ arises from BCH formula and represents higher-order commutators between $\slashed{\partial}$ and $\slashed{A}$. Effectively, the nonlocal interaction yields a tower of an infinite number of unconventional local interactions, involving the electromagnetic and spinor fields. At the perturbative level, once $\Lambda$ is taken to be a typical high-energy scale (for example, the Planck scale), this means that the dimensionless coupling constant $\alpha=\frac{e}{\Lambda}\ll1$; thereby, the nonlocal effects on the electromagnetic field are Planck-suppressed. In practical terms, each vertex of the Feynman diagrams will carry a factor proportional to $e^{-i\alpha\slashed{A}}$, which can be expanded in power series of $\alpha$. As a result, only the leading term will be relevant as a first approximation, while the other ones can be safely neglected. Therefore, in this environment, the nonlocal interacting Dirac equation, after some algebraic manipulation, looks like
\begin{eqnarray}
    \nonumber&\bigg[&\Box\, g(\Box,\slashed{\partial})+\left(ie(\partial_{\mu}A^{\mu}+A_{\mu}\partial^{\mu}\right)-e^2A_{\mu}A^{\mu})g(\Box,\slashed{\partial})+\\
    &+&\frac{e}{2}\sigma^{\mu\nu}F_{\mu\nu}\,g(\Box,\slashed{\partial})+...-m^2\bigg]\Psi=0,
\end{eqnarray}
where the first term in the above equation is just the free nonlocal Dirac equation, while the other terms represent the coupling of the electromagnetic field with the nonlocal Dirac operator. The ellipses stand for corrections in the dimensionless parameter $\alpha$. We plan to study the quantum aspects of this model elsewhere.

\subsection{Nonrelativistic limit and nonlocal Pauli equation}

In the local approach of the Dirac equation, the nonrelativistic limit is reached by assuming a gauge field weakly coupled to fermions, thereby preventing them from attaining relativistic velocities. We now investigate how that limit applies to the nonlocal Dirac equation with minimal electromagnetic coupling. To do that, let us first examine Eq.\eqref{equ2} in the momentum space,
\begin{equation}
    \bigg[\bigg(\slashed{p}-e\slashed{A}\bigg)f(\slashed{p}_{\Lambda}-\alpha\slashed{A})-m\bigg]\tilde{\Psi}(p)=0,
    \label{nonre}
\end{equation}
where $\slashed{p}_{\Lambda}\equiv \dfrac{\slashed{p}}{\Lambda}$. Before proceeding further, let us turn our attention to the form factor $f(\slashed{p}_{\Lambda}-\alpha\slashed{A})$. Note that $\alpha\ll 1$ in the weak-field limit, as previously discussed. Since in this subsection we investigate nonlocality effects in the nonrelativistic regime, we can therefore safely set $\alpha=0$ in the last equation.

At this stage, we make some choices. First, we use $f(\slashed{p}_{\Lambda})=e^{-\frac{\slashed{p}}{\Lambda}}$, which corresponds to the Fourier transform of the form factor \eqref{IIi}. Then, setting $\alpha=0$ in Eq.\eqref{nonre} yields
\begin{equation}
    \bigg[\bigg(\slashed{p}-e\slashed{A}\bigg)e^{-\frac{\slashed{p}}{\Lambda}}-m\bigg]\tilde{\Psi}(p)=0.
    \label{kkk}
\end{equation}
To proceed, it is necessary to explicitly specify the representation of the gamma matrices and other physical quantities. We adopt the standard Dirac representation for the gamma matrices, that is,
\begin{align}
\gamma^0 &= 
\begin{pmatrix}
I & 0 \\[6pt]
0   & -I
\end{pmatrix},
\quad\vec{\gamma} = 
\begin{pmatrix}
0         & \vec{\sigma} \\[6pt]
-\vec{\sigma} & 0
\end{pmatrix}.
\end{align}
Using Eq.\eqref{fa} the form factor can be cast into the following matrix form,
\begin{align}
e^{-\frac{\slashed{p}}{\Lambda}} &= 
\begin{pmatrix}
\cosh(\frac{mc}{\Lambda})-\frac{E}{mc^2}\sinh(\frac{mc}{\Lambda}) & -\frac{1}{mc}\sinh(\frac{mc}{\Lambda})\vec{\sigma}\cdot\vec{p} \\[6pt]
\frac{1}{mc}\sinh(\frac{mc}{\Lambda})\vec{\sigma}\cdot\vec{p}   & \cosh(\frac{mc}{\Lambda})+\frac{E}{mc^2}\sinh(\frac{mc}{\Lambda})
\end{pmatrix}.
\end{align}
Note that, for the sake of convenience, we have restored the speed of light $c$ in the above and subsequent equations 
Furthermore, assuming the spinor
\begin{equation}
    \tilde{\Psi}=\begin{pmatrix}
\chi \\[6pt]
\xi 
\end{pmatrix},
\end{equation}
and the vector potential $A_{\mu}=\left(\dfrac{\phi}{c},\vec{A}\right)$, Eq. \eqref{kkk} reduces to a couple of equations for two-component spinors $\chi$ and $\xi$, namely,
\begin{eqnarray}
    \left(a_{11}b_{11}-a_{12}b_{12}-mc\right)\chi+\left(a_{11}b_{12}+a_{12}b_{22}\right)\xi&=&0;\label{91}\\    \left(a_{12}b_{11}-a_{11}b_{12}\right)\chi+\left(a_{12}b_{12}+a_{11}b_{22}+mc\right)\xi&=&0,\label{92}
\end{eqnarray} 
where we have defined
\begin{equation}
\begin{split}
    a_{11}&=\left(\frac{E}{c}-\frac{e\phi}{c}\right) ;\\
    a_{12}&=\vec{\sigma}\cdot\vec{\Pi};\\
    b_{11}&=\cosh\left(\frac{mc}{\Lambda}\right)-\frac{E}{mc^2}\sinh\left(\frac{mc}{\Lambda}\right);\\
    b_{12}&=-\frac{1}{mc}\sinh\left(\frac{mc}{\Lambda}\right)(\vec{\sigma}\cdot\vec{p});\\  
    b_{22}&=\cosh\left(\frac{mc}{\Lambda}\right)+\frac{E}{mc^2}\sinh\left(\frac{mc}{\Lambda}\right),
    \end{split}
    \label{coeff}
\end{equation}
with $\vec{\Pi}=\vec{p}-\frac{e}{c}\vec{A}$. 

Let us now take into account the non-relativistic limit, which corresponds to taking  $E\approx mc^2+E_c$, where $E_c\equiv \frac{|\vec{p}|^2}{2m}$ is the classical energy. In this case and considering the weak-field regime, Eq.\eqref{92} becomes 
\begin{equation}
    \xi=-\frac{1}{mc\left(1+e^{\frac{mc}{\Lambda}}\right)}\left((\vec{\sigma}\cdot \vec{\Pi})e^{-\frac{mc}{\Lambda}}+(\vec{\sigma}\cdot \vec{p})\sinh\left(\frac{mc}{\Lambda}\right)\right)\chi.
    \label{eps1}
\end{equation}
Substituting Eq.\eqref{eps1} into Eq.\eqref{91} and using the coefficients \eqref{coeff}, we arrive at
\begin{eqnarray}
   \nonumber &&\Bigg\{\left(\frac{E_c-e\phi}{c}\right)e^{-\frac{mc}{\Lambda}}+mc\left[e^{-\frac{mc}{\Lambda}}-1\right]+\frac{(\vec{\sigma}\cdot \vec{\Pi})(\vec{\sigma}\cdot \vec{p})}{mc}\sinh\left(\frac{mc}{\Lambda}\right)\Bigg\}\chi=\\
   \nonumber&=&\frac{1}{mc\left(1+e^{\frac{mc}{\Lambda}}\right)}\bigg\{(\vec{\sigma}\cdot \vec{\Pi})^2-(\vec{\sigma}\cdot \vec{p})^2\sinh^2\left(\frac{mc}{\Lambda}\right)+e^{\frac{mc}{\Lambda}}\sinh\left(\frac{mc}{\Lambda}\right)(\vec{\sigma}\cdot \vec{\Pi})(\vec{\sigma}\cdot \vec{p})-\\
   &-&e^{-\frac{mc}{\Lambda}}\sinh\left(\frac{mc}{\Lambda}\right)(\vec{\sigma}\cdot \vec{p})(\vec{\sigma}\cdot \vec{\Pi})\bigg\}\chi,
    \label{eps}
\end{eqnarray}

In order to simplify \eqref{eps}, we must make use of some vectorial identities:
\begin{eqnarray}
\nonumber(\vec{\sigma}\cdot \vec{\Pi})(\vec{\sigma}\cdot\vec{p})&=&\frac{1}{2}\left(\lbrace\sigma_i ,\sigma_j\rbrace+\left[\sigma_i ,\sigma_j\right]\right)\Pi_i p_j\\
\nonumber&=&\vec{\Pi}\cdot\vec{p}+i\vec{\sigma}\cdot\left(\vec{\Pi}\times \vec{p}\right)\\
&=&|\vec{\Pi}|^2+\frac{e}{c}\vec{\Pi}\cdot\vec{A}+i\frac{e}{c}\vec{\sigma}\cdot(\vec{A}\times\vec{p}) \label{spp};\\
    \nonumber(\vec{\sigma}\cdot \vec{\Pi})^2 &=& \frac{1}{2}\left(\lbrace\sigma_i ,\sigma_j\rbrace+\left[\sigma_i ,\sigma_j\right]\right)\Pi_i \Pi_j \\
    \nonumber &=& |\vec{\Pi}|^2+i\,\vec{\sigma}\cdot \left(\vec{\Pi}\times \vec{\Pi}\right)\\
    &=&|\vec{\Pi}|^2-i\frac{e}{c}\vec{\sigma}\cdot\vec{B},
    \label{sigK}
\end{eqnarray}
where we have used the identity
\begin{eqnarray}
    \nonumber\vec{\Pi}\times \vec{\Pi}&=&-\frac{e}{c}\left(\vec{p}\times\vec{A}+\vec{A}\times\vec{p}\right)\\
  \label{Pp}  &=&i\frac{e}{c}\vec{B}
\end{eqnarray}
in the above formulas.

Plugging Eqs.\eqref{spp} and \eqref{sigK} into  Eq.\eqref{eps} and after some algebraic manipulations, one finds 
 \begin{eqnarray}
     \nonumber E_c \chi&=&\Bigg\{e\phi-mc\left[1-e^{\frac{mc}{\Lambda}}\right]+\frac{1}{2m}\left(\frac{e^{\frac{mc}{2\Lambda}}}{\cosh(\frac{mc}{2\Lambda})}\right)|\vec{\Pi}|^2-\\
    \nonumber&-&\frac{e}{2mc}\left(\frac{e^{\frac{mc}{2\Lambda}}}{\cosh\left(\frac{mc}{2\Lambda}\right)}+\frac{\sinh\left(\frac{mc}{\Lambda}\right)}{\cosh(\frac{mc}{2\Lambda})e^{\frac{mc}{2\Lambda}}}\right)\vec{\sigma}\cdot\vec{B}-\frac{e^{\frac{mc}{2\Lambda}}\sinh^2 \left(\frac{mc}{\Lambda}\right)}{2m\cosh\left(\frac{mc}{2\Lambda}\right)}|\vec{p}|^2 -\\
    &-&\frac{\sinh\left(\frac{mc}{\Lambda}\right)}{mc}\left[c\vec{\Pi}\cdot\vec{p}-ie\vec{\sigma}\cdot(\vec{A}\times\vec{p})-\frac{e\sinh\left(\frac{mc}{\Lambda}\right)e^{-\frac{mc}{2\Lambda}}}{2\cosh\left(\frac{mc}{2\Lambda}\right)}(\vec{\Pi}\cdot\vec{A}-\vec{A}\cdot\vec{\Pi})\right]\Bigg\}\chi.
 \end{eqnarray}
 The former equation represents the nonlocal version of the Pauli equation. Taking $\Lambda\to \infty$ recovers the standard (local) Pauli equation. Note that this equation exhibits a nonlocal correction to the gyromagnetic ratio ($g$-factor) of the particle \cite{Keshavarzi:2021eqa}:
 \begin{eqnarray}
    g= 2\left(\frac{e^{\frac{mc}{2\Lambda}}}{\cosh\left(\frac{mc}{2\Lambda}\right)}+\frac{\sinh\left(\frac{mc}{\Lambda}\right)}{\cosh(\frac{mc}{2\Lambda})e^{\frac{mc}{2\Lambda}}}\right).
 \end{eqnarray}
To get more insight into this result, we expand $g$ in power series of $\frac{mc}{\Lambda}$ ( since the nonlocal scale satisfies $\Lambda\gg mc$)
\begin{equation}
    g=2+3\frac{mc}{\Lambda}+\mathcal{O}\left[\left(\frac{mc}{\Lambda}\right)^2\right].
\end{equation}
The local Pauli equation yields $g=2$; quantum fluctuations from QFT introduce small corrections, which in our scenario are due to the nonlocal effects.

\section{Summary and Conclusions}

In this work, we have formulated a spin-1/2 nonlocal theory in which the form factor is an entire function of the Dirac operator. The main idea behind proposing this theory is that, due to its spinorial structure, it is more natural to consider a form factor that depends on the Dirac operator instead of the d'Alembert operator.

We initially considered a free nonlocal spin-1/2 field theory by replacing the free Dirac operator $\slashed{\partial}$ with $\slashed{\partial} f(\slashed{\partial})$, where $f$ is an entire function. Next, we found the field equations and succeeded in rewriting them in terms of the spectrum of the square of the free Dirac operator. Furthermore, we demonstrated that each component of the Dirac spinor does not satisfy a nonlocal Klein-Gordon equation, as occurs in the local case. This happens because of the presence of an additional term proportional to the free Dirac operator, as shown in Eq. \eqref{nlde}. We studied free fermionic particles within this nonlocal model; in particular, an analytical expression for their dispersion relations was obtained for an exponential-like form factor. We showed that the dispersion relation approaches the standard result as nonlocal effects are suppressed (i.e., when $\Lambda\to\infty$), while it departs from the standard result as the nonlocal effects become relevant in the UV limit (i.e., as $\Lambda\sim p$, with $p$ representing large momenta). 

In the second part of this paper, we dealt with quantum corrections within the nonlocal spin-1/2 field theory in the presence of a Yukawa coupling. We computed the contributions for the fermionic one-loop effective action of this model using an approximate method. In this environment, we concluded that the one-loop effective action is highly suppressed by $\Lambda$ in the IR limit, making clear that the nonlocal effects are subleading in this regime. In contrast, in the UV limit, the nonlocal effects emerge in the one-loop effective action as leading terms. 

In the last part of this work, we minimally coupled a $U(1)$ gauge field to the nonlocal spin-1/2 field. We explicitly demonstrated that this theory is invariant under gauge transformations. Moreover, besides deriving the spinor field equation, we showed that the nonlocal interaction between the electromagnetic and spinor fields can be viewed as an infinite tower of unusual local interactions. As an application, we investigated its nonrelativistic limit and derived a nonlocal version of the Pauli equation. In this scenario, we found that the $g$-factor acquires contributions from nonlocal terms, which entails $(g-2)\neq 0$, thereby providing a nontrivial magnetic anomaly purely of nonlocal origin.     

The natural continuation of this study could consist in the study of the nonlocality approach within the spinors formalism in curved spaces with application in cosmology and also the implications of the parity and Lorentz symmetry violation in the context of gravitational waves, for example. We plan to address these issues in the forthcoming paper.
%%%%%%%%%%%%%%%%%%%%%%%%%%%%%%%%%%%%%%%%%%%%%%%%%%%%%%%%%%%%%%%%%%%%%%%%%%%%%%%%%%%%%%

{\bf Acknowledgments.} 
 %PJP would like to thank CAPES for financial support.
 This work was supported by Conselho Nacional de Desenvolvimento Cient\'{\i}fico e Tecnol\'{o}gico (CNPq) and Paraiba State Research Foundation (FAPESQ-PB). PJP would like to thank the Brazilian agency CNPq for financial support (PQ--2 grant, process 
 No. 307628/2022-1).  The work by AYP has been supported by the CNPq project No. 303777/2023-0. PJP, AYP and JRN  thank the Brazilian agency FAPESQ-PB for financial support (process No. 150891/2023-7). The authors also acknowledge financial support from the Spanish Grants  PID2020-116567GB-C21, PID2023-149560NB-C21 funded by MCIN/AEI
/10.13039/501100011033, and by CEX2023-001292-S funded by MCIU/AEI.  The paper is based upon work from COST Action CaLISTA CA21109 supported by COST (European Cooperation in Science and Technology).

%%%%%%%%%%%%%%%%%%%%%%%%%%%%%%%%%%%%%%%%%%%

\end{document}